\documentclass[preprint]{elsarticle}

\journal{Annals of Physics}

\begin{document}

\begin{frontmatter}

\title{Continuous and discontinuous realizations of first-order phase transitions}

\author{Matthias Hempel}
\address{4057 Basel, Switzerland}
\ead{matthias.hempel@freenet.de}

\begin{abstract}
First-order phase transitions are commonly associated with a discontinuous behavior of some of the thermodynamic variables and the presence of a latent heat. In the present study it is shown that this is not necessarily the case. Using standard thermodynamics, the general characteristics of phase transitions are investigated, considering an arbitrary number of conserved particle species and coexisting phases, and an arbitrary set of state variables. It is found that there exist two diﬀerent possible types of realizations of a phase transition. In the first type, one has the immediate replacement of a single phase with another one. As a consequence, some of the global extensive variables indeed behave \textit{discontinuously}. In the second type, one has instead the gradual (dis-) appearance of a single phase over a range of the state variables. This leads to a \textit{continuous} behavior of the (global) basic thermodynamic variables. Furthermore, in this case it is not possible to define a latent heat in a trivial manner. It is derived that the latter (former) case happens if the number of extensive state variables used is larger or equal (lower) than the number of coexisting phases. The choice of the state variables thus place a crucial role for the qualitative properties of the phase transition.
\end{abstract}


\end{frontmatter}

\section{Introduction}
\label{sec_intro}
First-order phase transitions play an important role in various diﬀerent areas
of physics and chemistry, but also in every-day life. The most prominent ex-
ample, which every human encounters almost every day, is the boiling of water.
Some other first-order phase transitions are the subject of present-day fundamental 
research. For example in relativistic heavy-ion collision experiments, e.g.,
done at CERN or at the future FAIR facility at GSI, physicists are searching
for signs of an anticipated phase transition in which hadrons are dissolved into
quarks, their elementary building blocks. This so-called QCD phase transition
can also occur in neutron stars, if the so-far unknown phase transition density
is low enough. The phase transition of water takes place at about 10 orders
of magnitude lower temperatures and 14 orders of magnitude lower densities
than the QCD phase transition. Nevertheless, both of these phase transitions
are described with the same thermodynamic framework for the coexistence of
phases.

The standard definition of a first-order phase transition, given, e.g., in Ref.~\cite{landau69}
is that one has spatially separated, coexisting phases which can be distinguished
from each other by some of their (local) thermodynamic quantities, e.g., by their
densities. Another definition is that the first derivatives of the thermodynamic
potential behave discontinuously, corresponding to the Ehrenfest classification.
In its original formulation \cite{ehrenfest33} (see also Ref.~\cite{jaeger98}), 
the Gibbs free energy is used
as the thermodynamic potential. In the present article it will be shown that
the choice of the thermodynamic potential, corresponding to a particular set of
natural state variables, plays a crucial role in this definition.\footnote{The original Ehrenfest classification has known deficits regarding second- and higher-
order phase transitions. Very often second-order phase transitions are actually characterized
by continuous, but divergent second derivatives of the thermodynamic potential ($\lambda$-transition),
as for example in the superfluid transition of $^4$He. For a thorough discussion of the Ehrenfest
classification and its historical evolution, see Ref.~\cite{jaeger98}.}

Another common definition is that a first-order phase transition involves a
`latent heat', which is only applicable for a phase transition at constant temperature. 
An example for a first-order phase transition at non-constant temperature 
is the boiling of water in a
pressure cooker. Here the boiling takes place over a range of temperatures, in
contrast to the constant boiling temperature of water in an open pot. As the temperature does not stay constant, it is not possible to define a latent heat in a simple manner.

This example illustrates the importance of the used state variables: for boil-
ing water in an open pot, the pressure is kept constant, in a pressure cooker
it is the volume, and this leads to very diﬀerent characteristics of the phase transition. 
In the present work these two variants
are referred
to as continuous and discontinuous realizations of a first-order phase transition. 
The term ‘realization’ is meant to refer to a particular thermodynamic process, corresponding 
to a particular set of state variables which are changed in a continuous
way.

Phase transitions can also change their characteristics if one goes to a multi-
component system, i.e., with several conserved particle species or quantum numbers. 
One example is the aforementioned QCD phase transition in neutron stars.
Neutron stars are described as being at zero temperature and in hydrostatic
equilibrium, so that the pressure is decreasing strictly monotonically with radius. 
The conserved quantum numbers are baryon number and the net electric
charge, which has to be zero to maintain charge neutrality. Usually local charge
neutrality is considered, and then the phase transition happens at a single radius 
inside the star. If instead the electric charge is treated as a second globally
conserved quantum number (ignoring any Coulomb interactions, which is a commonly used simplification in this research field), a spatially extended phase-coexistence region occurs, as the phase transition takes place over
a range of pressures \cite{glendenning92}. In the field of neutron star physics, these two diﬀerent
realizations are known as `Maxwell' and `Gibbs' phase transitions, respectively.
Eﬀects of additionally conserved quantum numbers are also discussed in the
context of heavy-ion collisions, see, e.g., Ref.~\cite{greiner87}. In chemistry and chemical
physics, a phase transition with phase coexistence of two (or more) macroscopic
phases with diﬀerent chemical compositions is also known as a `non-congruent'
phase transition, see the IUPAC definition \cite{clark94} and Refs.~\cite{iosilevskiy03,iosilevskiy10,hempel13}.

In the present paper, the most general case of a phase transition of a multi-
component substance with any number of conserved particle species and any
number of coexisting phases (respecting Gibbs phase rule) is considered. It
will be shown that only the number of extensive state variables used, $\cal E$, and
the number of involved phases, $\cal K$, determine whether the realization of the
phase transition is continuous (${\cal E} \geq {\cal K}$), where all basic thermodynamic variables
behave continuously, or discontinuous (${\cal E} < {\cal K}$)
where some of the extensive thermodynamic variables behave discontinuously. 
Furthermore, in the former case, the intensive
variables such as temperature or pressure are usually changing during the phase
transition and therefore it is not possible to define a latent heat. In the latter case the phase transition is happening at a single value
of the intensive variables and the concept of a `latent heat' can be applied. The dimensionality of phase
coexistence regions in phase diagrams also just depend on the number of intensive and
extensive state variables that are used.

Throughout the paper complete thermodynamic equilibrium is assumed.
Meta-stable or unstable states and aspects of nucleation are thus not of rel-
evance for the present study. Furthermore, finite size eﬀects are not considered,
which means that the diﬀerent phases which appear are always treated in the
thermodynamic limit.

The article is structured as follows: In Sec.~\ref{sec_vars_total}, the basic thermodynamic
variables und relations are introduced which are used throughout the paper.
In Sec.~\ref{sec_coex}, the conditions for phase coexistence and the consequences for the
dimensionality and dependency of phase coexistence regions in terms of the
state variables are analyzed. Sec.~\ref{sec_boundary} discusses the same aspects, but for the
boundaries of the phase coexistence regions. In Sec.~\ref{sec_proppt}, it is shown that there
are only two types of realizations of first-order phase transitions possible, the
continuous phase (dis-) appearance and the discontinuous phase replacement.
It is also shown that it is just the number of extensive state variables and the number
of coexisting phases which determine the type of realization. In Sec.~\ref{sec_latentheat}, the
latent heat and change of internal energy are analyzed for the two types, and it
is found that a latent heat is only involved in the discontinuous realization. To
illustrate the found results, Sec.~\ref{sec_example} presents some example phase diagrams and
example realizations of a phase transition for a substance with a triple point.
Sec.~\ref{sec_otherclass} discusses the implications of the found results in the context of other classifications of
phase transitions. Sec.~\ref{sec_summary} gives a summary and the main conclusions.

\section{Basic thermodynamic variables and relations}
\label{sec_vars_total}
This section introduces the basic thermodynamic variables and standard thermodynamic relations used in the present study. 
A thermodynamic system with $\cal N$ globally conserved particle species $k=1,..., {\cal N}$ is considered.\footnote{Other globally conserved quantities can be included in the same way, but for brevity the
following discussion is restricted to particle species. With ‘global conservation’ it is meant
that if there are several phases in equilibrium, the particles can be exchanged among the
diﬀerent phases, and only the total particle numbers $N_k$ of the entire system are conserved.} In addition to the particle numbers $N_k$, there are the volume $V$ and the entropy $S$ as extensive (also called additive) thermodynamic variables. In total there are thus ${\cal N}+2$ global extensive variables. Furthermore, there are the ${\cal N}+2$ conjugate intensive (also called non-additive) variables: the ${\cal N}$ chemical potentials $\mu_k$, the pressure $P$, and the temperature $T$.\footnote{The term `basic thermodynamic variables' used in the introduction and the abstract refers to these  $2({\cal N}+2)$  variables.} In the following, the extensive variables are denoted as $X_l$ and the intensive variables as $Y_l$, with $l=1, ..., {\cal N}+2$. The internal energy, which is also extensive, is then given as 
\begin{equation}
E=\sum_{l=1}^{{\cal N}+2}Y_l X_l \; .
\end{equation}

Next, a particular set of ${\cal N}+2$ independent, natural state variables is chosen, where from each of the ${\cal N}+2$ conjugate pairs either the extensive or the intensive variable is being selected. Such a set of natural state variables completely defines the state of the system. This is always true for a single phase, exceptions for particular states of phase coexistence are discussed below. After a possible renumbering, w.l.o.g., let $\mathbf{\tilde  X}=(\tilde X_i) = (X_i)$, $i=1, ..., \cal{E}$ and $\mathbf{\tilde Y}=(\tilde Y_j)=(Y_j)$, $j={\cal E}+1, ..., {\cal N}+2$ denote the tuples of the $\cal{E}$ chosen independent extensive and the $\cal{I}$ chosen independent intensive state variables, respectively, where 
\begin{equation}
{\cal E} + {\cal I} = {\cal N}+2 \, . \label{eq_easy}
\end{equation}
The tilde is introduced to distinguish the state variables from the full set of thermodynamic variables.
Because of the Gibbs-Duhem relation, not all of the ${\cal N}+2$ intensive variables are independent, which implies ${\cal I} < {\cal N}+2$, or equivalently ${\cal E}\geq 1$. This can also be understood intuitively: at least one extensive state variable has to be used to fix the size of the system. 
By selecting the state variables there are ${\cal E}$ dependent intensive and ${\cal I}$ dependent extensive variables, $Y_i$ and $X_j$, respectively, remaining. 

The choice of the state variables also fixes the natural thermodynamic potential $\Phi(\mathbf{\tilde X},\mathbf{\tilde Y})$, which is obtained from the internal energy $E$ by Legendre transformations: 
\begin{eqnarray}
 \Phi(\mathbf{\tilde X},\mathbf{\tilde Y})&=&E-\sum_{j={\cal E}+1}^{{\cal N}+2}  \tilde Y_j X_j \label{eq_pot0}\\
                          &=&\sum_{i=1}^{{\cal E}} Y_i(\mathbf{\tilde X},\mathbf{\tilde Y}) \tilde X_i \; . \label{eq_pot}
\end{eqnarray}
The dependent variables obey
\begin{eqnarray}
Y_i &=&  \frac{\partial \Phi}{\partial \tilde X_i} \; , \label{eos_tildeyi} \\
X_j &=& -\frac{\partial \Phi}{\partial \tilde Y_j} \label{eos_tildexj} \; .
\end{eqnarray}
Note that this generalized notation can lead to opposite signs of thermodynamic quantities compared to their usual definition. This would for example be the case for the pressure, but does not have any effect on the following derivation.

\section{Phase coexistence}
\label{sec_coex}
The aim of this section is to derive general properties of phase coexistence regions and their dimensionality in the space of the state variables $\tilde X_i$ and $\tilde Y_j$. It is assumed that one has a state where $\cal K$ different phases $\kappa=1, ..., \cal K$ are in equilibrium. 
Each phase has its own thermodynamic potential ${\Phi}^\kappa$, internal energy $E^\kappa$, and thermodynamic variables $X^\kappa_l$, $Y^\kappa_l$. 
All relations of Sec.~\ref{sec_vars_total} also apply for these quantities of the single phases. However, because none of them will be used as state variables, tildes are omitted for the variables of the single phases in the following. Considering all extensive and intensive variables of the single phases and the total system, one has in total $2({\cal N}+2)({\cal K}+1)$ variables or degrees of freedom: $X_l, X_l^\kappa, Y_l, Y_l^\kappa$. In the following, step by step, all conditions are considered which reduce or constrain these degrees of freedom, to derive the dimensionality of the phase coexistence region.

The Gibbs' conditions for phase equilibrium among the $\cal K$ phases are:
\begin{eqnarray}
  Y_l=Y^{\kappa}_l \; . \label{eq_gibbs}
\end{eqnarray}
These conditions follow from the minimization of the total thermodynamic potential and express chemical, thermal, and mechanical equilibrium. They imply that the intensive variables $Y^{\kappa}_l$ can be omitted, so that one is left with the $({\cal N}+2)({\cal K}+2)$ degrees of freedom $X_l, X_l^\kappa$, and $Y_l$. 
Next it is used that the total value of each extensive quantity $X_l$ is given by the sum over all phases:
\begin{eqnarray}
X_l&=&\sum_{\kappa=1}^{\cal K} X_l^\kappa  \label{eq_ext} \; ,
\end{eqnarray}
leaving $({\cal N}+2)({\cal K}+1)$ degrees of freedom, which can, e.g., be chosen to be $X_l^\kappa$, and $Y_l$.

For each phase there are in total ${\cal N}+2$ equations of state of type (\ref{eos_tildeyi}) and (\ref{eos_tildexj}). To make use of them, it is helpful to realize  that it is possible to choose a different number of extensive and intensive state variables for the single phases than for the entire system. Here the equations of state are used in the special form where $\Phi^\kappa(X_1^\kappa,\mathbf Y^\kappa)=X_1^\kappa Y_1^\kappa$ is chosen as the thermodynamic potential with $\mathbf Y^\kappa=(Y_m^\kappa)$, $m=2, ..., {\cal N}+2$. This means for each phase one uses only one extensive state variable, which fixes the size of this phase. This simplifies the following derivation. The equations of state then read:
\begin{eqnarray}
Y_1^\kappa &=&  \frac{\partial \Phi^\kappa}{\partial X_1^\kappa} \label{eos1}\; , \\
X_m^\kappa &=& -\frac{\partial \Phi^\kappa}{\partial Y_m^\kappa} \label{eos2} \; .
\end{eqnarray}
This formulation has the advantage that one can use the Gibbs-Duhem relation
for each single phase - the intensive variables and ratios of extensive
variables of the single phases do not depend on their size $X_1^\kappa $:
\begin{eqnarray}
Y_1^\kappa &=&  Y_1^\kappa(\mathbf Y^\kappa)\label{gd1}  \; , \\
X_m^\kappa &=& -x_m^\kappa(\mathbf Y^\kappa)X_1^\kappa \label{gd2} \; ,
\end{eqnarray}
where $x_m^\kappa:=X_m^\kappa/X_1^\kappa$. In the following it is assumed that all phases have a non-vanishing size, i.e., $X_1^\kappa>0$. The case where some of the phases have zero size is discussed in Sec.~\ref{sec_boundary}.
Eqs.~(\ref{gd1}) represent  $\cal K$ additional constraints for the intensive variables, and Eqs.~(\ref{gd2}) represent ${\cal K}({\cal N}+1)$ additional constraints involving both intensive and extensive variables.\footnote{It has to be noted that the set of Eqs.~(\ref{eq_gibbs}), (\ref{eq_ext}), (\ref{eos1}), and (\ref{eos2}) automatically imply that the relations (\ref{eos_tildeyi}) and (\ref{eos_tildexj}) of the total system are fulfilled, i.e., these latter two equations do not carry additional information. Furthermore, one also gets $E=\sum_{\kappa=1}^{\cal K} E^\kappa$ and $\Phi=\sum_{\kappa=1}^{\cal K} \Phi^\kappa$, as it has to be.}

In total one is left with exactly ${\cal N}+2$ degrees of freedom. This means that the number of degrees of freedom of the $\cal K$-phase coexistence matches the number of state variables which are specified externally. By fixing the state variables, then all other variables can be determined. For a given state in the $\cal K$-phase coexistence region, it is in general possible to vary the state variables in an arbitrary direction without leaving the $\cal K$-phase coexistence region (its boundaries will be determined later). This implies that the phase coexistence region has a dimensionality $({\cal N}+2)$ in the $({\cal N}+2)$-dimensional space of state variables $\{\tilde X_i,\tilde Y_j\}$.

However, there is one important aspect in the derivation which has to be reexamined, by looking just at all the constraints for the intensive variables. Eqs.~(\ref{eq_gibbs}) and (\ref{gd1}) involve only intensive variables and are completely independent of the extensive variables. These two sets of equations leave ${\cal N}+2-{\cal K}$ intensive variables as degrees of freedom. Using the fixed values of the $\cal I$ intensive state variables $\tilde Y_j$ leaves ${\cal N}+2-{\cal K}-{\cal I}={\cal E}-{\cal K}$ (using Eq.~(\ref{eq_easy})) degrees of freedom. For ${\cal E} \leq {\cal K}$ it follows that the dependent intensive variables do not depend on the extensive state variables, $Y_i = Y_i(\tilde \mathbf Y)$, whereas for ${\cal E} > {\cal K}$ one has $Y_i = Y_i(\tilde \mathbf X, \tilde \mathbf Y)$.
For ${\cal E} < {\cal K}$ the intensive state variables are overdetermined. It implies the $\cal K$-phase coexistence has only a reduced dimension ${\cal I}+{\cal E}-{\cal K}<{\cal I}$ in the space of intensive state variables $\{\tilde Y_j\}$ compared to the full dimension $\cal I$. 

As a side aspect, from these results one can identify an interesting property of the case where one has the maximum possible number of phases in coexistence, ${\cal K}={\cal N}+2$ (Gibbs' phase rule). By considering ${\cal E}={\cal N}+2 \Leftrightarrow {\cal I} =0 $, so that ${\cal E} = {\cal K}$, it follows that all the intensive variables are fixed just by the equilibrium conditions and that they do not depend on the state variables, $Y_l=Y_l()={\rm const}$. 

Next, the consequences of ${\cal E} < {\cal K}$ for just the ${\cal K}({\cal N}+2)$ extensive variables $X_l^\kappa$ are discussed. As pointed out above, the dependent intensive variables $Y_i$ are independent of the extensive variables in this case. For fixed intensive state variables all other intensive variables are fixed too, and Eqs.~(\ref{gd2}) then represent ${\cal K}({\cal N}+1)$ constraints for the extensive variables $X_l^\kappa$, leaving $\cal K$ degrees of freedom. Using relations (\ref{eq_ext}) and the fixed values of the extensive state variables $\tilde X_i$, leaves ${\cal K} - {\cal E} > 0$ free variables. This means that the extensive variables $X_j$ and $X_1^\kappa$ remain undetermined, in the same way as the intensive variables are overdetermined. 

However, this does not affect the dimensionality of the phase coexistence region in the space of extensive state variables $\{\tilde X_i\}$: For fixed intensive state variables, the phase coexistence region still has the full dimension $\cal E$ in the space of extensive state variables. In total, the dimensionality of the phase coexistence region in the space $\{\tilde X_i, \tilde Y_j\}$ is thus reduced to ${\cal N}+2 +{\cal 
E}-{\cal K} < {\cal N} +2$. The results of this section are summarized in Table~\ref{table_propmix}.

\begin{table}
\begin{center}
\begin{tabular}{c c c c}
\hline
\hline 
case & dimensionality in $\{\tilde X_i,\tilde Y_j\}$ & \multicolumn{2}{c}{dependent variables}\\
\hline
$\cal K < \cal E$ & ${\cal N}+2$ & $ Y_i= Y_i(\mathbf{\tilde X}, \mathbf{\tilde Y})$&  $ X_j= X_j(\mathbf{\tilde X}, \mathbf{\tilde Y})$  \\
$\cal K = \cal E$ & ${\cal N}+2$ & $ Y_i= Y_i(\mathbf{\tilde Y})$ &  $ X_j= X_j(\mathbf{\tilde X}, \mathbf{\tilde Y})$  \\
$\cal E<\cal K$ & ${\cal N}+2+{\cal E}-{\cal K}$& $ Y_i= Y_i(\mathbf{\tilde Y})$ &  $X_j$ undetermined  \\
\hline
\hline
\end{tabular}
\end{center}
\caption{Properties of $\cal K$-phase coexistence regions. The dimension in the space of state variables $\tilde X_i,\tilde Y_j$ is given in the second column. The third and fourth column show the dependency of the dependent intensive and extensive variables, $Y_i$ and $X_j$, respectively, on the state variables. In the case $\cal E < \cal K$ the intensive state variables $\tilde Y_j$ are overdetermined, resulting in a reduced dimensionality. Furthermore, in this case the sizes of the phases and thus also the dependent extensive variables $X_j$ remain undetermined.}
\label{table_propmix}
\end{table}

\section{Boundaries of phase coexistence regions}
\label{sec_boundary}
To identify the boundary C of a certain phase coexistence region A, one can use the additional constraints that exactly $\cal L$, $1\leq {\cal L} (<{\cal K})$, of the phases of A have zero size. W.l.o.g.\ and a possible renumbering it is assumed that the last ${\cal L}$ phases are vanishing on C: $X_1^{\lambda} = 0$, $\lambda= {\cal K}-{\cal L}+1, ..., {\cal K}$, and $X_1^{\kappa'} \neq 0$, $\kappa'= 1, ..., {\cal K}-{\cal L}$. 

First the case ${\cal K} \leq {\cal E}$ is analyzed. This means the phase coexistence region A has a dimension ${\cal N} +2$ in the space of the state variables $\{\tilde X_i, \tilde Y_j \}$, see Table~\ref{table_propmix}. Because of the $\cal L$ additional constraints $X_1^{\lambda} = 0$, C has a reduced dimension 
\mbox{${\cal N}+2-{\cal L}$.}
Next, the boundary C of a phase coexistence region A with ${\cal E}<{\cal K}$ is analyzed. As discussed in Sec.~\ref{sec_coex}, in this case the extensive variables $X_j$ remain undetermined, the intensive state variables $\tilde Y_j$ are overdetermined, and the intensive variables $Y_i$ are independent of the extensive state variables, see Table~\ref{table_propmix}. This implies that the additional constraints $X_1^{\lambda} = 0$ do not have any effect on the intensive variables. 
However, they can affect the extensive variables. Using
the same argumentation as in Sec.~\ref{sec_coex}, it is found that for ${\cal K}>{\cal L}+{\cal E}$ the dependent extensive variables $X_j$ remain undetermined and the dimensionality in the space of state variables $\{\tilde X_i, \tilde Y_j\}$ is still ${\cal N}+2+{\cal E}-{\cal K}$, as for the phase coexistence region A. The only effect of the vanishing phases is that they possibly can reduce the extension of C in the extensive state variables $\tilde X_i$. If ${\cal K}={\cal L}+{\cal E}$, the dimensionality is also still ${\cal N}+2+{\cal E}-{\cal K}={\cal N}+2-{\cal L}$, but now the additional constraints are sufficient to determine the dependent extensive variables $X_j$. Again, C has the same dimension as A, but a reduced extension in the extensive state variables. 
If ${\cal K}<{\cal L}+{\cal E}$, for fixed intensive variables there are more constraints on the extensive variables than degrees of freedom, so that there is an overdetermination of the extensive state variables $\tilde X_i$. As a consequence, in comparison to A the dimensionality of C is reduced from ${\cal N}+2+{\cal E}-{\cal K}$ to ${\cal N}+2-{\cal L}$. 
These results are summarized in Table~\ref{table_boundary}.

\begin{table}
\begin{center}
\begin{tabular}{c c c c }
\hline
\hline 
case & dimensionality in $\{\tilde X_i,\tilde Y_j\}$ & \multicolumn{2}{c}{dependent variables}\\
\hline
$\cal K < \cal E$ & ${\cal N}+2-{\cal L}$ & $ Y_i= Y_i(\mathbf{\tilde X}, \mathbf{\tilde Y})$&  $ X_j= X_j(\mathbf{\tilde X}, \mathbf{\tilde Y})$  \\
$\cal K = \cal E$ & ${\cal N}+2-{\cal L}$ & $ Y_i= Y_i(\mathbf{\tilde Y})$ &  $ X_j= X_j(\mathbf{\tilde X}, \mathbf{\tilde Y})$  \\
$\cal E < {\cal K}< {\cal E}+{\cal L}$ & ${\cal N}+2-{\cal L}$& $ Y_i= Y_i(\mathbf{\tilde Y})$ &  $ X_j= X_j(\mathbf{\tilde X}, \mathbf{\tilde Y})$   \\
${\cal K}= {\cal E}+{\cal L}$ & ${\cal N}+2+{\cal E}-{\cal K}$& $ Y_i= Y_i(\mathbf{\tilde Y})$ &  $ X_j= X_j(\mathbf{\tilde X}, \mathbf{\tilde Y})$  \\
${\cal E}+{\cal L}<{\cal K}$ & ${\cal N}+2+{\cal E}-{\cal K}$& $ Y_i= Y_i(\mathbf{\tilde Y})$ &  $X_j$ undetermined  \\
\hline
\hline
\end{tabular}
\end{center}
\caption{Properties of the boundary of a $\cal K$-phase coexistence region where exactly $\cal L$ phases have zero size. The columns of the table are the same as in Table~\ref{table_propmix}.}
\label{table_boundary}
\end{table}

For the following discussion it is important to realize that the solution for the phase coexistence on C would not change, if some or all of the phases $\lambda$ with vanishing size were not part of the phase equilibrium at all. Thus the states on C also correspond to phase equilibria of the non-vanishing phases $\kappa'$ together with an arbitrary selection of the vanishing phases $\lambda$. 

\section{Possible realizations of phase transitions and their properties}
\label{sec_proppt}
This section gives a classification of diﬀerent realizations of first-order phase transitions and their characteristic properties. `Realization' is meant to refer to a given path $\Gamma: (\mathbf{\tilde X}(r),\mathbf{\tilde Y}(r))$ through the space of the state variables $\{\tilde X_i,\tilde Y_j\}$, described by the parameter $r$ on which the state variables are varied continuously. The path is supposed to start within a particular phase coexistence region A and to end within a different ``neighboring'' phase coexistence region B. The cases where A and/or B actually just consist of a single phase is also included. 
The phase coexistence regions A and B have to have a dimension ${\cal N}+2$ in the space $\{\tilde X_i,\tilde Y_j\}$. Otherwise, some of the state variables had to be tuned to exact values to reach them, which is excluded in the following. Therefore one has ${\cal K}^A\leq {\cal E}$ and ${\cal K}^B\leq {\cal E}$ (see Table~\ref{table_propmix}) and w.l.o.g.\ one can assume ${\cal K}^B \leq {\cal K}^A \leq {\cal E}$. The transition from A to B, that occurs at $r^{t}$ with $(\mathbf{\tilde X}(r^{t}),\mathbf{\tilde Y}(r^{t}))=:(\mathbf{\tilde X}^{t},\mathbf{\tilde Y}^{t})$, is supposed to lie inside the ${\cal N}+1$-dimensional boundary C between A and B. This also implies that no other phase coexistence regions than A and B are connected to C at the transition point. Because C has a dimension ${\cal N}+1$, there are only two options for the properties of C, as can be inferred from Tables~\ref{table_propmix} and \ref{table_boundary}:
\begin{enumerate}
 \item ${\cal K}^C \leq {\cal E}$ and ${\cal L} =1$.

As A and B should be the only coexistence regions connected to C, there is just one possible solution: C is the boundary of A where exactly one of the phases of A has zero size, and B is the phase coexistence region consisting of all of the other phases. Thus one has ${\cal K}^C={\cal K}^A$ and ${\cal K}^B = {\cal K}^A-1$. 

 \item ${\cal K}^C ={\cal E}+1$ and ${\cal L} \leq 1$.

C corresponds to (a part of) a phase coexistence region of ${\cal E}+1$ phases. 
$\cal L$ can either be 0 (corresponding to Table~\ref{table_propmix}) or 1 (corresponding to Table~\ref{table_boundary}). It was noted in Sec.~\ref{sec_boundary} before, that setting one of the ${\cal K}^C$ phases to zero size (i.e.~$\cal L=1$), does not change the
dimensionality of the phase coexistence region, but potentially just reduces
its extension in the extensive state variables. In contrast, C would have a lower dimensionality if $\cal L\geq 2$. This means one has ${\cal K}^C = {\cal E} + 1$ phases on C, with either arbitrary or undetermined sizes, or with exactly one phase of zero size. From this it follows, that A and B can have just exactly one phase less than C, i.e.~${\cal K}^A = {\cal K}^B = {\cal E}$, whereas A and B consist of two different subsets of the phases of C.

\end{enumerate}

Next it is discussed how the dependent intensive and extensive variables, $Y_i$ and $X_j$, respectively, behave at the transition point in the two different cases.
\begin{enumerate}
 \item On C, A and B have identical solutions for all variables as the additional
phase of A has zero size. This implies that all dependent intensive and extensive variables, $Y_i$ and $X_j$, respectively, behave continuously across the transition point $(\mathbf{\tilde X}^{t},\mathbf{\tilde Y}^{t})$.

 \item Because A and B differ by one phase and all of their phases keep a finite size within the vicinity of C by construction (otherwise one would have ${\cal L} \geq 2$ on C), A and B have different solutions for $X_j$ at $(\mathbf{\tilde X}^{t},\mathbf{\tilde Y}^{t})$. The dependent extensive variables $X_j$ therefore behave discontinuously across the transition. On C, i.e., on the transition point itself, the dependent extensive variables $X_j$ remain undetermined as noted before. 

In Sec.~\ref{sec_boundary} it was noted that for ${\cal E} < {\cal K}^C (= {\cal E}+1)$ the dependent intensive variables $Y_i$ are not affected from setting some of the phases of C to zero size. Therefore A, B and C all have to have the same, identical solution for the the dependent intensive variables $Y_i$. This implies that the dependent intensive variables $Y_i$ behave continuously across the transition. 

\end{enumerate}

These results give a complete picture of the phase diagram and the types of phase transitions that are possible without fine tuning of the state variables, i.e., between coexistence regions A and B of dimension ${\cal N} + 2 $ that are separated
from each other by a ${\cal N} + 1$-dimensional boundary C. There are just two types of realization:
\begin{enumerate}
\item \textbf{Continuous phase (dis-) appearance}: One has  ${\cal K}^B + 1= {\cal K}^A \leq {\cal E}$, and exactly one of the phases of A disappears in a continuous manner at
the transition point situated on C. B is made of all the remaining phases.
\item \textbf{Discontinuous phase replacement}: One has ${\cal K}^A = {\cal K}^B = {\cal E}$, A and B differ in exactly one phase, and all ${\cal E}+1$ phases of A and B coexist on C. In the crossing of the boundary between two different sets of $\cal E$ phases, 
one phase is replaced discontinuously with another one.
\end{enumerate}

\section{Latent heat and change of internal energy}
\label{sec_latentheat}
Next it is investigated whether or not the two diﬀerent realizations of first-
order phase transitions involve a latent heat $\Delta Q^{t}$. Latent heat is defined as the energy released or absorbed during a constant-temperature process, other than mechanical work or by transfer of matter. It was shown in Sec.~\ref{sec_proppt} that all intensive variables variables, including the temperature, behave continuously at the transition point from A to B and thus the transition always represents an isothermal process. The heat transfered to the system in the transition from A to B is given as 
\begin{eqnarray}
 \Delta Q^{t}&=&T^{t} \Delta S^{t} \\
       &=&T^t(S^B(\mathbf{\tilde X}^{t},\mathbf{\tilde Y}^{t})-S^A(\mathbf{\tilde X}^{t},\mathbf{\tilde Y}^{t})) \; ,
\end{eqnarray}
where $S^A$, respectively $S^B$, denotes the total entropy of phase coexistence region A, respectively B, and $T^ t$ the common temperature at the transition point.
In the continuous phase (dis-) appearance, all extensive variables are continuous, which leads to $\Delta Q^{t}=0$. For a discontinuous phase replacement in general one has $\Delta Q^{t} \neq 0$, unless the entropy is part of the state variables and thus behaves continuously by construction.

It is also instructive to consider the change of the internal energy of the system at the transition point:
\begin{equation}
\Delta E^{t} = E^B(\mathbf{\tilde X}^{t},\mathbf{\tilde Y}^{t})-E^A(\mathbf{\tilde X}^{t},\mathbf{\tilde Y}^{t}) \; ,
\end{equation}
The internal energy is related to the thermodynamic potential, Eq.~(\ref{eq_pot0}), by:
\begin{eqnarray}
E(\mathbf{\tilde X},\mathbf{\tilde Y})=\Phi(\mathbf{\tilde X},\mathbf{\tilde Y})+\sum_{j={\cal E}+1}^{{\cal N}+2}  \tilde Y_j X_j \; .
\end{eqnarray}
The thermodynamic potentials of the two phase coexistence regions are always equal at the transition point because the dependent intensive variables $Y_i$ behave continuously, and the extensive state variables by construction, too (cf.\ Eq.~(\ref{eq_pot})). Thus one gets:
\begin{equation}
\Delta E^{t}=\sum_{j={\cal E}+1}^{{\cal N}+2}  \tilde Y_j \left(X_j^B(\mathbf{\tilde X}^{t},\mathbf{\tilde Y}^{t})-X_j^A(\mathbf{\tilde X}^{t},\mathbf{\tilde Y}^{t})\right)\; .
\end{equation}
In continuous phase (dis-) appearance this becomes zero and in discontinuous
phase replacement one has $\Delta E^t \neq 0$, because of the (dis-) continuity of the dependent extensive variables. Only phase replacement can lead to the release or absorption of latent heat and a discontinuous change of the internal energy at the transition point. 

\section{Example realizations of phase transitions of a one-component substance with a triple point}
\label{sec_example}
To illustrate the general results, in the following examples of diﬀerent realizations of phase transitions of a one-component substance (${\cal N}=1$) are discussed. It is considered that the substance exists in three phases: gas, liquid, and solid, similarly as ordinary water. \footnote{Water has the well known anomaly that the solid ice phase has a lower density than the
liquid phase. This leads to a negative slope of the solid-liquid coexistence line in the pressure-
temperature phase diagram. In the present study, a substance without such an anomaly is
considered. For a realistic phase diagram of water, see Ref.~\cite{glasser04}; for a thorough discussion of
diﬀerent types of phase diagrams of a two-component substance, see §97 of Ref.~\cite{landau69}.}
A certain region around the triple point, where all three phases are in coexistence, is investigated. Only the dimensionality of the phase coexistence regions and the qualitative properties of different realizations are discussed. The scales, quantitative properties, and slopes of phase boundaries are not of relevance here.

\begin{figure}
\includegraphics[width=0.5\columnwidth]{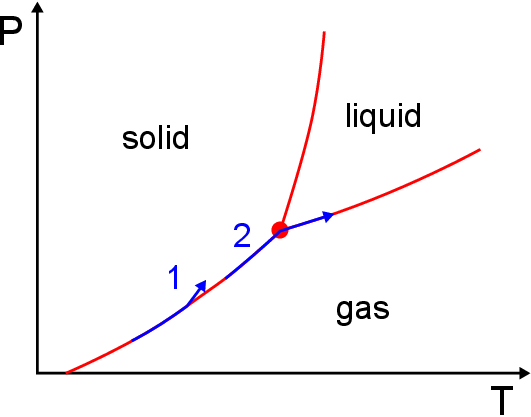}
\caption{\label{fig_1}An example phase diagram for a one-component substance, with constant number of particles $N$, and the pressure $P$ and the temperature $T$ as the independent state variables. The red lines correspond to phase coexistence of two phases. The red dot marks a triple point, where all three phases are in equilibrium. If the red lines are crossed in a thermodynamic
process of the aforementioned state variables, this results in a discontinuous realization of the phase transition. The blue arrows show two examples of realizations where the state variables of Fig.~\ref{fig_2} are used, see text.}
\end{figure}

In the first example a fixed number of particles $N$ in a container of variable size inside a heatbath is considered. It is assumed that the pressure and temperature of the heatbath can be controlled and varied freely. The other two state variables for the substance in the container besides $N$ are thus pressure~$P$ and temperature $T$, and therefore ${\cal E}=1$. Because $N$ is assumed to be constant, it is sufficient to plot the phase diagram for the remaining two state variables, as is done in Fig.~\ref{fig_1}. There exist the three areas of the single phases, separated from each other by three different coexistence lines where always two phases are in coexistence. These lines intersect in the triple point, where all three phases coexist. The dimensionality of the single phases in $\{T,P\}$ is 2, of the two-phase coexistence lines it is 1, and of the triple point it is 0. Note that either $T$
or $P$ have to be fine-tuned to precise values to hit the triple point in a phase
transition using these state variables. Therefore the only relevant realizations
of phase transitions are those that cross a two-phase coexistence line. They are discontinuous phase replacements because ${\cal K}=2 > 1 ={\cal E}$. In such a realization the system jumps from one single phase to another one.
 
\begin{figure}
\includegraphics[width=0.5\columnwidth]{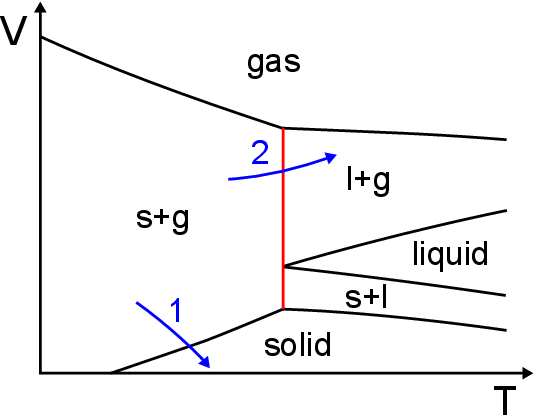}
\caption{\label{fig_2}The phase diagram for a constant number of particles $N$, and for the volume $V$ and the temperature $T$ as the independent state variables. ``s'' denotes the solid, ``l'' the liquid and ``g'' the gas phase. The red line is the triple line, on which all three phases are in equilibrium. Its crossing results in a discontinuous realization. The blue arrows show examples of a continuous phase disappearance (1) and a discontinuous phase replacement (2).}
\end{figure}

The arrows which are drawn in Fig.~\ref{fig_1} belong to realizations with a different set of state variables namely in which the pressure is replaced by the volume $V$, so that ${\cal E}=2$. This could be done, e.g., with a particle container of adjustable size inside the heat bath. The corresponding phase diagram is shown in Fig.~\ref{fig_2}. Compared to Fig.~\ref{fig_1}, the three two-phase coexistence lines change to two-dimensional areas in this plot. The triple-point becomes a triple line at constant temperature, that is extended in $V$. At the triple line, the three two-phase coexistence areas meet.

Next, two illustrative examples for a continuous and a discontinuous realization are discussed in more detail. In realization 1, one starts with a phase coexistence of the solid and the gas, whereas the volume fractions of the two phases are determined by the state variables. By increasing the temperature and decreasing the volume, as shown in the plot, the system approaches the pure solid phase. Gradually the solid transforms into gas until it disappears completely and only the gas phase is left at the transition point. This realization is a continuous phase disappearance where all variables behave continuously. 
 
In realization 2, one starts with a similar state of phase coexistence of the solid and the gas. But this time the system is heated and compressed in such a way, that it ends in phase coexistence of the liquid and the gas. The triple line is crossed, on which one has ${\cal K}=3 > 2 ={\cal E}$. By approaching the transition point, the volumes of the solid and the gas phase approach some particular (non-vanishing) values. When the transition point has been passed, the solid is replaced by the liquid and the volumes of the two respective phases jump to new values. On the transition point all three phases coexist, whereas their volume fractions remain undetermined. The extensive variables that are not used as state variables behave discontinuously at the transition point. This is
an example for discontinuous phase replacement.

It is interesting to study the paths of the discussed realizations in the $\{T,p\}$-plane, which is shown in Fig.~\ref{fig_1}. One sees that the intensive variables $T$ and $P$ change continuously, both for the continuous phase dis-appearance and the discontinuous phase replacement.

\begin{figure}
\includegraphics[width=0.5\columnwidth]{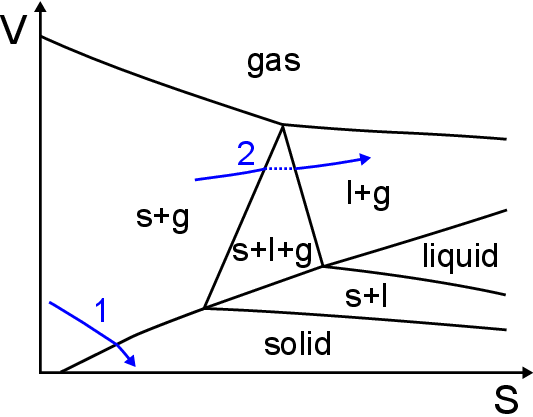}
\caption{\label{fig_3}The phase diagram for a constant number of particles $N$, and for the volume $V$ and the entropy $S$ as the independent state variables. There are no discontinuous phase replacements for this set of state variables. The blue arrows belong to the realizations shown in Fig.~\ref{fig_2}.}
\end{figure}

For ${\cal E}=3$, the volume $V$, the number of particles $N$ and the entropy $S$ are used as state variables. It could be realized, e.g., by an isolated container that can be heated in a controlled way and that is of adjustable size. The corresponding phase diagram is shown in Fig.~\ref{fig_3}. The following phase coexistence regions exist which all are two-dimensional in $\{S,V\}$ and three-dimensional in $\{S,V,N\}$: There are the three different single phases. Between each pair of them a two-phase coexistence region exists. Between the three phase coexistence regions, there is the triple-phase coexistence region with ${\cal K}=3$. Because ${\cal E}=3$, all possible realizations are continuous. With this set of state variables one could realize a three-phase coexistence without fine tuning.

It is instructive to investigate the example realizations of Fig.~\ref{fig_2} in the parameter space of the extensive variables of Fig.~\ref{fig_3}. Now it becomes obvious, why the two examples are continuous and discontinuous. The entropy $S$ is the only variable which can behave discontinuously for the set of state variables of Fig.~\ref{fig_2}. In example 2, the temperature is changed continuously. Thus the entire triple phase region in Fig.~\ref{fig_3} is crossed at once, because the three phases can coexist only at a single value of the temperature. This is visualized by the dotted line in Fig.~\ref{fig_3}. In example 1 there is no discontinuous change of entropy.

\section{Relation to other classifications of phase transitions}
\label{sec_otherclass}
In this section the connection between the present results and other classifications of phase transitions, in particular the Ehrenfest classification, is discussed. First of all it is noted that in the present study coexistence of two or more \textit{spatially separated} phases was considered. The different phases can be distinguished from each other, e.g.~by some of the ``densities'' $x_m^\kappa$. According to Landau \cite{landau69}, such a situation corresponds to a ``phase transition of the first kind''.

Next, the Ehrenfest classification of phase transitions is considered. This classification usually refers to the Gibbs free energy $G(N,T,P)$ of a one component system, ${\cal N}=1$, in which only one extensive state variable, namely the particle number $N$ is used. Thus it refers to a specific set of state variables with ${\cal E}=1$.
According to the Ehrenfest classification a phase transition is of first order, if at least one of the first derivatives of the Gibbs free energy is discontinuous, and of second order if all first derivatives are continuous, but some of the second are not.\footnote{Note that in his original formulation \cite{ehrenfest33}, Ehrenfest did not treat the particle number $N$ as a free variables, and only the derivatives with respect to $T$ and $P$ were considered.} If one applies the results from the present investigation to this set of state variables‚ $(N,T,P)$, because ${\cal E}=1<{\cal K}\geq 2 $, all realizations or possible phase transitions are discontinuous phase replacements. The $X_j$ (entropy $S$ and volume $V$ in this case), which are the first derivatives of the Gibbs free energy, behave discontinuously. This means that all realizations with this set of state variables would be classified as being first order phase transitions according to the Ehrenfest classification. In general, for a discontinuous phase replacement the Ehrenfest classification always gives a first-order phase transition: Because ${\cal E} < {\cal K} \leq {\cal N}+2$, one has ${\cal I}\geq 1$, i.e.~there is always at least one intensive state variable. The first derivative of the thermodynamic potential w.r.t.~an intensive state variable is an extensive variable, which behaves discontinuous, as was shown above.

However, for other choices of the state variables and the corresponding thermodynamic potential it is possible to obtain a second or higher order phase transition according to Ehrenfest, which is shown in the following example. If only extensive state variables are used $(N, S, V)$, even for the maximum number of phases in coexistence, ${\cal K} =3$, one has $\cal K = \cal E $ and therefore all realizations are continuous phase (dis-) appearance. The thermodynamic potential for this choice of the state variables is the internal energy. Its first derivatives are the intensive variables $\mu$, $T$ and $P$. For continuous phase (dis-) appearance these behave continuously and therefore such a realization would be second (or higher) order according to Ehrenfest. Obviously, this is not only the case for this particular example of state variables, but for all possible realizations of continuous phase (dis-) appearance.

\section{Summary and conclusions}
\label{sec_summary}
The main topic of this investigation are different realizations of first order phase transitions, considering the general case of any number of coexisting phases and conserved particle numbers. 'Realization' refers to a continuous change of a particular set of state variables. Without fine tuning of the state variables, just two characteristic types of realizations are possible. The first
type is continuous phase (dis-) appearance and the second type discontinuous
phase replacement.  It was shown that one has the first type if, and only if, the number of extensive state variables ${\cal E}$ is larger or equal to the total number of phases ${\cal K}$ involved in the realization.

In continuous phase (dis-) appearance with $\cal K \leq \cal E$, a single phase is appearing or disappearing at the transition point in a continuous way. At the transition point the volume of this phase is zero, and it is continuously increasing afterwards, respectively continuously decreasing before. All basic thermodynamic variables behave continuously.\footnote{ ``basic thermodynamic variables'' refers to the intensive and global extensive variables. Ratios
of extensive variables of the individual phases instead always behave discontinuously in a first-order phase transition by definition.} 
As a consequence, there is no latent heat associated with this realization of a first-order phase transition and in the Ehrenfest classification such a realization would be classified to be of second (or higher) order.

In a discontinuous phase replacement, one has phase coexistence of ${\cal E}$ phases before and after the transition, whereas one of the phases is replaced by another one in a discontinuous manner at the transition point, i.e. one has ${\cal K} = {\cal E} +1$ in total.\footnote{Phase coexistence regions with ${\cal K} > {\cal E} +1$ would require fine tuning of the state variables which was ignored in the present article.} 
At the transition point, the volumes of all phases change discontinuously, leading to a discontinuous behavior of the dependent extensive variables, i.e., those extensive variables which are not part of the state variables. As a consequence, there is a latent heat related to this type of realization, unless the entropy is part of the state variables. At the transition point itself, the volumes of the single phases and the dependent extensive variables remain
undetermined. In the Ehrenfest classification such a realization would be classified to be of first order.

It has to be noted that the intensive variables behave continuously in both
types of realizations. In addition to these general conclusions about the two
types of possible realizations of first-order phase transitions, the dimensionality of phase coexistence regions and their boundaries, and the dependencies of the dependent variables were derived, and it was shown that these properties also depend only on the number of extensive state variables and the number of coexisting phases. These results are summarized in Tables \ref{table_propmix} and \ref{table_boundary}.

A simple understanding of the important role of the choice of the state variables is given in the following. On the one hand, during phase coexistence all intensive variables of all phases have to be equal. On the other hand, each intensive state variable that is used represents a direct constraint for the intensive variables. If one has ${\cal N} + 2 < {\cal I} + {\cal K}  \Leftrightarrow {\cal E} < {\cal K}$, this leads to an overdetermination of the intensive variables, even without using any information from the extensive variables. The dependent extensive variables in turn remain undetermined. In contrast, for ${\cal E} \geq  {\cal K}$, there is no such over- and undetermination. All dependent variables can be determined, and all state variables have to be used to do so. For ${\cal E} = {\cal K}$ one just has the special situation that the dependent intensive variables are determined by the intensive state variables alone. One could say that the intensive state variables carry more relevant or constraining information for the phase coexistence than the extensive state variables. 

\section{Acknowledgement}
I thank J\"urgen Schaﬀner-Bielich for the very useful discussions and comments during an early stage of the work presented in this article.

\bibliography{literat}

\end{document}